\documentclass[a4paper,prb,showpacs,twocolumn]{revtex4}

\usepackage{amsfonts}
\usepackage{graphicx}
\usepackage{color}
\usepackage{amsmath}
\usepackage{amssymb}
\usepackage{latexsym}
\usepackage{psfrag}

\begin{document}
\title{Fabry-P\'{e}rot and Aharonov-Bohm interference in ideal graphene nanoribbons}
\author{S. Ihnatsenka}
\affiliation{Department of Science and Technology, Link\"{o}ping University, SE-60174, Norrk\"{o}ping, Sweden}
\email{sergey.ignatenko@liu.se}

\begin{abstract}
Quantum-mechanical calculations of electron magneto-transport in ideal graphene nanoribbons are presented. In noninteracting theory, it is predicted that an ideal ribbon that is attached to wide leads should reveal Fabry-P\'{e}rot conductance oscillations in magnetic field. In the theory with Coulomb interaction taken into account, the oscillation pattern should rather be determined by the Aharonov-Bohm interference effect. Both of these theories predict the formation of quasi-bound states, albeit of different structures, inside the ribbon because of strong electron scattering on the interfaces between the connecting ribbon and the leads. Conductance oscillations are a result of resonant backscattering via these quasi-bound states.
\end{abstract}
\maketitle

\section{Introduction}

Interference is a fundamental phenomenon in which two coherent waves superpose to form a resultant wave of greater or lower amplitude. Electrons, as particles that share wave properties, reveal quantum interference, which is a counterpart of classical interference in which the wave-function interferes with itself --- a phenomenon that was demonstrated in the double-slit experiments in the 1960s.\cite{Jon61-Don73} Electron quantum interferometers have been used to explore many physical phenomenon including, for example, anyon statistics in fractional quantum Hall effect,\cite{Nak20} and the interplay between charge and wave electron properties.\cite{Zha09} Nearly two decades ago, graphene emerged as a two-dimensional (2D) material with unusual electronic properties that were best described by theories for massless relativistic particles.\cite{Cas09} These properties have been exploited in graphene-based interferometers.\cite{Dep21-Ron21, Mak18, She08-Mor15-Rus08, Wei17, Kra20-Han17, Mil19, You09, Ric13, Oks14} Fabry-P\'{e}rot and Aharonov-Bohm interference in a perpendicular magnetic field has recently been demonstrated in graphene interferometers with high visibility and less obscuration by Coulomb charging effects.\cite{Dep21-Ron21} The aim of this study is to provide the quantum-mechanical theory of electron interference in one of the simplest graphene-based devices --- graphene nanoribbon (GNR) --- that realistically represents an experimental setup in which a ribbon is connected to two wide graphene electrodes,\cite{Bis15} and which might be used further as a building block for more complex interferometer structures. 

In optics, Fabry-P\'{e}rot (FP) interference occurs in a system of two parallel surfaces when optical waves are only allowed to pass through if they are in resonance with the active region of the device. An electron analog to an optical FP interferometer has been demonstrated in many experimental realizations.\cite{Lia01, Cam07, Dep21-Ron21, Kra20-Han17, You09, Geh16, Ric13, Oks14} Similarly to an optical wave, for an electron wave to pass through the active region, multiple reflections between two transmission barriers are required to be in phase, a condition referred to as constructive interference. This condition can be written as
\begin{equation} \label{eq:deBroglie0}
i\frac{\lambda_F}{2}=l, \quad i=1,2, \dots ,
\end{equation}
where $\lambda_F$ is the de Broglie wavelength of an electron at the Fermi energy $E_F$, and $l$ is the length of the active region. Having the condition \eqref{eq:deBroglie0} satisfied, electron is resonantly (nearly perfect) transmitted or reflected through a device. In the system that is studied in this manuscript, a pair of interfaces between narrow and wide graphene regions serve as electronic mirrors that confine electron waves in analogy to confinement of light in an optical Fabry-P\'{e}rot cavity.

A mesoscopic device, in which electrons can be guided into spatially separated paths when placed in perpendicular magnetic field, reveals another physical phenomenon --- the Aharonov-Bohm (AB) interference.\cite{Datta} Conductance through this device shows a peak each time a flux enclosed by the area separating electron paths, $S$, changes by the flux quantum $\phi_0=h/e$. An associated phase shift of electron wave function in the paths accumulates a value of $2\pi$. As a function of magnetic field $B$, conductance shows periodic oscillations with period 
\begin{equation} \label{eq:AB}
\Delta B = \frac{\phi_0}{S}.
\end{equation}
Graphene AB interferometers have been demonstrated in Refs. \onlinecite{She08-Mor15-Rus08, Mil19}. In the literature, an interferometer operating on AB effect is also referred to as Mach-Zehnder\cite{Ji03, Wei17, Mak18} or FP\cite{Dep21-Ron21} interferometer, depending on device configuration and path counting details. However, for GNR configuration considered here, the difference between FP and AB interference is important, leading to different observables, and so these terms are opposed here. Comment on AB terminology will be given later on.

Theoretical studies of electron quantum interference in graphene have focused on $p$-$n$ junctions,\cite{Shy08, Zeb18, Mak18, Cre07, Ric13, Kra20-Han17, Wei17, Oks14} quantum rings,\cite{Wur10, Mre16} quantum antidots\cite{Mil19} and GNRs.\cite{Ngu19, Dar09, Roc10, Geh16} For $p$-$n$-$p$ junction, Shytov \textit{et. al.}\cite{Shy08} showed that FP interference depends on the electron's incident angle and, in a small perpendicular magnetic field, scattering amplitude changes sign, which both are signatures of Klein tunneling through a graphene $p$-$n$ junction. This was later confirmed experimentally.\cite{You09} 
In Ref. \onlinecite{Ric13}, Klein tunneling was shown to yield a strong collimation in transmission and FP resonances to occur in bipolar and unipolar regimes due to reflection at internal $n$-$p$ ($p$-$n$) interfaces or at the outer contacts.  
Electron trajectories can be further modified by placing a graphene layer onto hexagonal boron nitride, whose lattice slightly mismatches graphene lattice.\cite{Kra20-Han17} This allows us to engineer a superlattice with subtle features in a FP interference pattern. At high magnetic fields, when current carrying one-dimensional edge channels form,\cite{Datta} the graphene $p$-$n$ junction starts hosting these edge channels along its interface and acts as an AB interferometer. The edge channels propagate, separated from each other by a distance, and are coupled at end points, where the $p$-$n$ junction meets physical graphene boundary, thus forming an encircled area.\cite{Wei17, Mak18} In Ref. \onlinecite{Mak18}, making a specific assumption about charge density profile along the interface, quantum-mechanical simulation has provided a clear characterization of the regime where the AB effect takes place and showed how it can be distinguished from classical snake states, all findings supported by comparison with experimental data. Similarly, AB conductance oscillations were predicted for a system of a locally induced $p$-$n$ junction (e.g. by the tip of an atomic force microscope).\cite{Zeb18} If currents along the $p$-$n$ junction, or GNR physical edge, propagate for a sufficiently long distance and equilibrate, then the fractional plateaus of quantum Hall conductance are predicted to additionally appear.\cite{Zeb18} For quantum rings made of graphene, numerical calculations show conventional AB magnetoconductance oscillations (similarly to 2D GaAs-based electron gases) when magnetic fields are not high enough to bring the system into the quantum Hall regime.\cite{Wur10} In the quantum Hall regime, a graphene antidot reveals AB interference that is either dominated by electron interaction or single-particle resonant tunneling depending on the coupling to the antidot bound edge channels.\cite{Mil19} AB conductance oscillations might also be revealed in GNR placed on a stepped substrate, such that $B$ spatially varies, and oppositely propagating edge states are obtained in terrace and facet zones of the step, resulting in inter-edge scattering.\cite{Ngu19} At zero $B$, Darancet \textit{et. al.}\cite{Dar09} demonstrated the existence of FP interference oscillations in GNRs in analogy with optics.

The object of this theoretical study is a GNR that is similar to GNRs that are typically fabricated in experiments based on nanolithography.\cite{Bis15} In these experiments, 2D graphene is first covered by protective resist layer. The desired GNR geometry is then patterned by electron-beam lithography, followed by plasma etching. The resulting structure exposes the hexagonal graphene lattice (usually very defective) along its physical boundaries. It should be noted that long narrow ribbon, wide semi-infinite leads and the interfaces between them are all integrated parts of the system that contribute to the measured signal and, for accurate analysis, need to be treated as a whole. This is conceptually similar to an extended molecule in molecular electronics\cite{Kir07} and is in line with ideas about the role of contacts in ballistic conduction.\cite{Datta, Lan57} Accounting for ribbon-to-lead interfaces has shown, for example, to result in transport gaps that are very different from predictions based on simple theories that approximate the system only by its narrowest part (i.e., GNR to have the straight geometry with ribbon and leads of the same width).\cite{Ihn21} Consequently, this study considers electronic and transport properties of an "extended" GNR that is otherwise ideal, free of defects. 

In this manuscript, energy and transport properties of ideal GNRs \textit{in perpendicular magnetic field} are studied by quantum-mechanical calculations in noninteracting and interacting approaches. The latter accounts for the long-range repulsive Coulomb interaction between charged particles within the Hartree approximation. The noninteracting approach shows that an ideal ribbon attached to two wide leads shows Fabry-P\'{e}rot conductance oscillations in the magnetic field. This is a result of resonant backscattering on quasi-bound states that are formed inside the ribbon. Conductance oscillations in the noninteracting approach follow the interference condition \eqref{eq:deBroglie0}. The inclusion of Coulomb interaction into theory brings qualitatively new physics with conductance oscillations determined rather by Aharonov-Bohm effect, where magnetic field period is determined by the ribbon area, Eq. \eqref{eq:AB}. The effects of FP and AB interference are unexpected (from a naive viewpoint) for this kind of structure because it is inherently open, atomically ideal and contains no potential barriers. Observing either of these effects allows one to discriminate between whether or not Coulomb interaction dominates electron transport in GNR.

This manuscript is organized as follows. In the first part, electron magneto-transport is studied in a simpler noninteracting approach, and electronic wave interference in ideal GNR is discussed. In the second part, the effects due to inclusion of Coulomb interaction into the theory are presented for the same device geometry. Differences in the results of the noninteracting model are then analyzed. Finally, the implications of a possible experiment are discussed, followed by the conclusion. The theoretical model is presented briefly here; for detailed formulation of the model and the computational method, the reader is referred to the earlier publications in Refs. \onlinecite{Ihn12, Ihn21}.

The model is based on the tight-binding Hamiltonian in the Hartree approximation\cite{Fer07, Ihn12, Ihn21, Are10}
\begin{align} \label{eq:1}
H &= -\sum_{\langle i,j \rangle} t_{ij} a_i^{\dagger}a_j + \sum_i V_i^H a_i^{\dagger} a_i \\
V_i^H &= \frac{e^2}{4\pi\varepsilon_0\varepsilon} \sum_{j\neq i} n_j \left(\frac{1}{| \mathbf{r}_i - \mathbf{r}_j |} - \frac{1}{\sqrt{| \mathbf{r}_i - \mathbf{r}_j |^2 +4b^2}} \right) %\label{eq:VH}
\end{align}
where $a_i^{\dagger}$ ($a_i$) is the creation (destruction) operator of the electron on the site $i$ and the angle brackets denote the nearest neighbour indices. The magnetic field is included via Peierls substitution $t_{ij}=-t\exp(i\frac{2\pi}{\phi_0}\int_{\mathbf{r}_i}^{\mathbf{r}_j}\mathbf{A}\cdot\mathbf{dr})$, where $t=2.7$ eV. The Hartree potential $V_i^H$ describes the long-range Coulomb interaction of electron at $i$-th atom with uncompensated charge density $-en$ in the system;\cite{Ihn12, Ihn21, Fer07, Sil08, Are10} $-en_j$ is the electron charge at the lattice site $j$ and $\mathbf{r}_j$ is the position vector of that site; $\varepsilon$ (=3.9 of SiO$_2$ for results below) is the dielectric permittivity; $b$ (=10 nm) is the distance to the screening gate electrode. If $V_i^H=0$, then the resulting Hamiltonian becomes the standard noninteracting approximation for electrons on a graphene lattice.\cite{Rei02, Cas09} In order to solve the transport problem within this tight-binding model, the recursive Green's function method is used, together with the self-consistent solution for the electrostatic potential and charge density.\cite{Ihn12, Ihn21, Datta}

The system studied is armchair GNR with armchair edges along straight segments of ribbon and semi-infinite leads. It is assumed that there are no defects in bulk or at the edges. The central part of the GNR has width $w=10$ nm and is connected to twice wider leads via mesoscopically smooth junctions,\cite{Ihn21} see the outlines in the top panels in Figs. \ref{fig:1} and \ref{fig:2}.  

\section{Fabry-P\'{e}rot interference in the noninteracting model}

%*********************************************************
\begin{figure*}[ht]
\includegraphics[keepaspectratio,width=1.8\columnwidth]{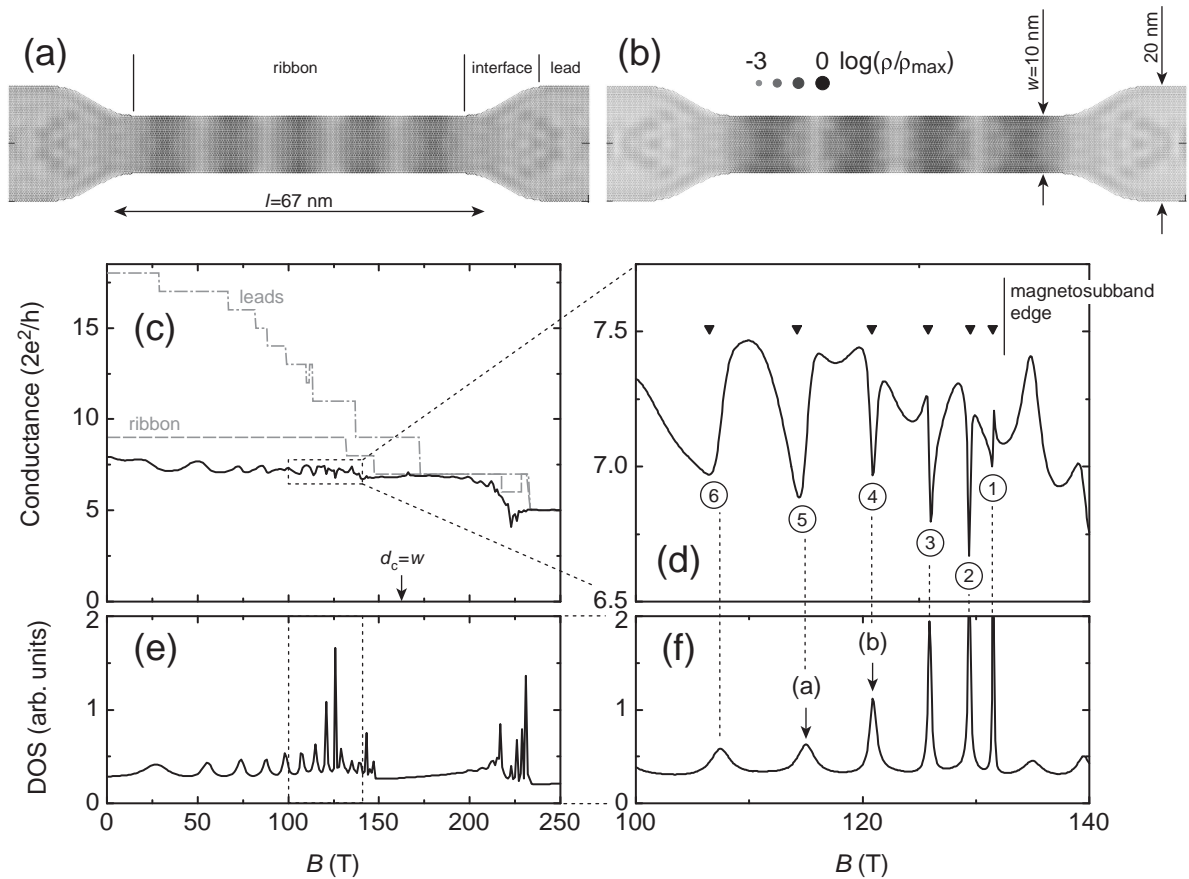}
\caption{(c),(d) Conductance and (e),(f) density of states (DOS) in GNR as a function of magnetic field calculated in the noninteracting approach. (d) and (f) show the magnified areas where the resonances due to Fabry-P\'{e}rot interference are designated by matching conductance dips with enhanced DOS. The integers in (d) mark resonances, of which fifth and fourth are further detailed in local DOS plots (a) and (b). Local DOS, $\rho$, is grey-scale and dot-size coded in (a) and (b). The gray dashed and dash-dotted lines in (c) and (d) show conductance in the ribbon (the long narrow part sandwiched between the interfaces) and leads, respectively. The triangles in (d) mark $B$ corresponding to the resonant condition \eqref{eq:deBroglie0} counted from magnetosubband edge at $B=132.5$ T. The arrow in (c) marks $B$, for which the cyclotron diameter equals the ribbon width, $d_c=w$. $E_F=0.3$ $t$. Temperature $T=0$ K.}
\label{fig:1}
\end{figure*}
%********************************************************* 

The noninteracting model predicts that in a perpendicular magnetic field, $B$, both two-terminal conductance and density of states (DOS) in GNR oscillate with features being recognizable as a result of Fabry-P\'{e}rot interference, Fig. \ref{fig:1}. At $B=0$ T, conductance $G$ in GNR is determined by the electron scattering at the interfaces connecting the ribbon to the leads, which is very similar to strong scattering in graphene constrictions, where poor conductance quantization was found.\cite{Mun06, Ihn12} These wide-to-narrow interfaces serve as transmission barriers and intrinsic scattering sources, even though they are mesoscopically smooth, because of the broken graphene sublattice symmetry\cite{Cas09} along interface edges and multiple alternating zigzag and armchair terminations.\cite{Wur09-Lib16} At $B=0$ T, the inequality $G<G^{\text{ribbon}}<G^{\text{leads}}$ holds, where the ribbon superscript denotes the central part of GNR; see Fig. \ref{fig:1}(c).

From low to moderate $B$, the Landau levels start to develop, which affects the electron quantization subbands that are gradually pushed up in energy and depopulate.\cite{Datta} This can be observed as staircase decrease of $G^{\text{ribbon}}$ and $G^{\text{leads}}$ in Fig. \ref{fig:1}(c). As an electron subband moves up, the longitudinal part of electron energy and associated wavevector $k_{\parallel}$ gradually decreases. This gradual decrease brings the GNR system through a series of sharp resonances, which are pronounced as coupled $G$ dips and DOS peaks in Fig. \ref{fig:1}(c)-(f). Enhanced DOS implies constructive interference of the single-particle state inside the ribbon, while $G$ dip strongly suggests \textit{resonant backscattering} of the incident electron in the Bloch state of the semi-infinite lead on that state. $G$ oscillates with nearly constant amplitude of half of the conductance quanta $2e^2/h$ and with period scaling inversely with $B$. This large amplitude points to a nearly perfect resonant reflection for the highest occupied quantization subband. To get further physical insight, let us use the relationship between the wave vector of Dirac electrons and $B$: 
\begin{equation} \label{eq:kB}
k=\sqrt{\frac{2eB}{\hbar}},
\end{equation}
and then relate it to the de Broglie wavelength in \eqref{eq:deBroglie0} by $k=k_{\parallel}=2\pi/\lambda_F$. Counting from magnetosubband edge at $B=132.5$ T, the resonance condition \eqref{eq:deBroglie0} is satisfied for magnetic fields marked by the triangle pointers in Fig. \ref{fig:1}(d), if $l=67$ nm is used as a fit parameter. This $l$ value matches well the effective geometrical length of the ribbon, sandwiched between two interfaces, as indicated in Fig. \ref{fig:1}(a), and it clearly points out that \textit{the resonances in the noninteracting theory are due to Fabry-P\'{e}rot interference}. The integers $i$ in \eqref{eq:deBroglie0}, which are referred to as FP modes, are numbered in Fig. \ref{fig:1}(c). Meanwhile, $i=4$ and $i=5$ modes are selected as examples and their local DOS $\rho$ is shown in Fig. \ref{fig:1}(a),(b), where the node structure of the longitudinal electron wave is clearly discernible. Thus, the open-ended ideal GNR supports longitudinal resonant electron states in a magnetic field. This is similar to the effects found for carbon nanotubes\cite{Lia01} and ballistic constrictions in conventional 2D electron gas\cite{Kir89-Wan00} and GNRs\cite{Dar09} at zero $B$. 

The resonance $i=1$ at $B=131.5$ T in Fig. \ref{fig:1}(d) reveals a zigzag-like feature that is similar to a Fano-type resonance.\cite{Fan61} For $i=1$, electron wave is node-less, half of de Broglie wavelength, and the single-particle state has the strongest localization in comparison to other ($i>1$) states, defined by the width of DOS peak, see Fig. \ref{fig:1}(f). This is a realization of the Fano model in the sense that it includes the quasi-bound state, localized inside the ribbon, that interacts with continuum of extended states in the leads. When the energy of incident electron coincides exactly with resonance energy, the FP phase flips by $\pi$.\cite{Fan61} Note that similar anti-resonance has been found in meny mesoscopic devices, for example, a quantum antidot-based interferometer in conventional 2D electron gas\cite{Ihn09-AD} and short GNR at zero $B$.\cite{Geh16}

The conductance curves in Figs. \ref{fig:1}(c) and (d) show additional, superimposed oscillatory dependence. This is a result of interference from other magneto-subbands, which all provide propagating single-particle states at $E_F$ in the ribbon.

At high $B$, the Landau levels dominate the energy spectrum, and the edge channels are well defined and propagate along the physical boundaries of GNR without, or with little, backscattering.\cite{Datta} This denotes the quantum Hall regime and $G\approx G^{\text{ribbon}}\approx G^{\text{leads}}$ with little deviation at the transition regions between quantization plateaux. This can loosely be characterized by $B$ for which the cyclotron diameter $d_c=2 E_F /v_F eB$ is less than the ribbon width $w$, see the mark in Fig. \ref{fig:1}(c); $v_F=10^6$ m/s. Because the scattering efficiency of the wide-narrow interfaces is reduced, the Fabry-P\'{e}rot interference is hardly observable at high magnetic fields.

\section{Aharonov-Bohm interference in the Hartree model}

%*********************************************************
\begin{figure*}[th]
\includegraphics[keepaspectratio,width=1.8\columnwidth]{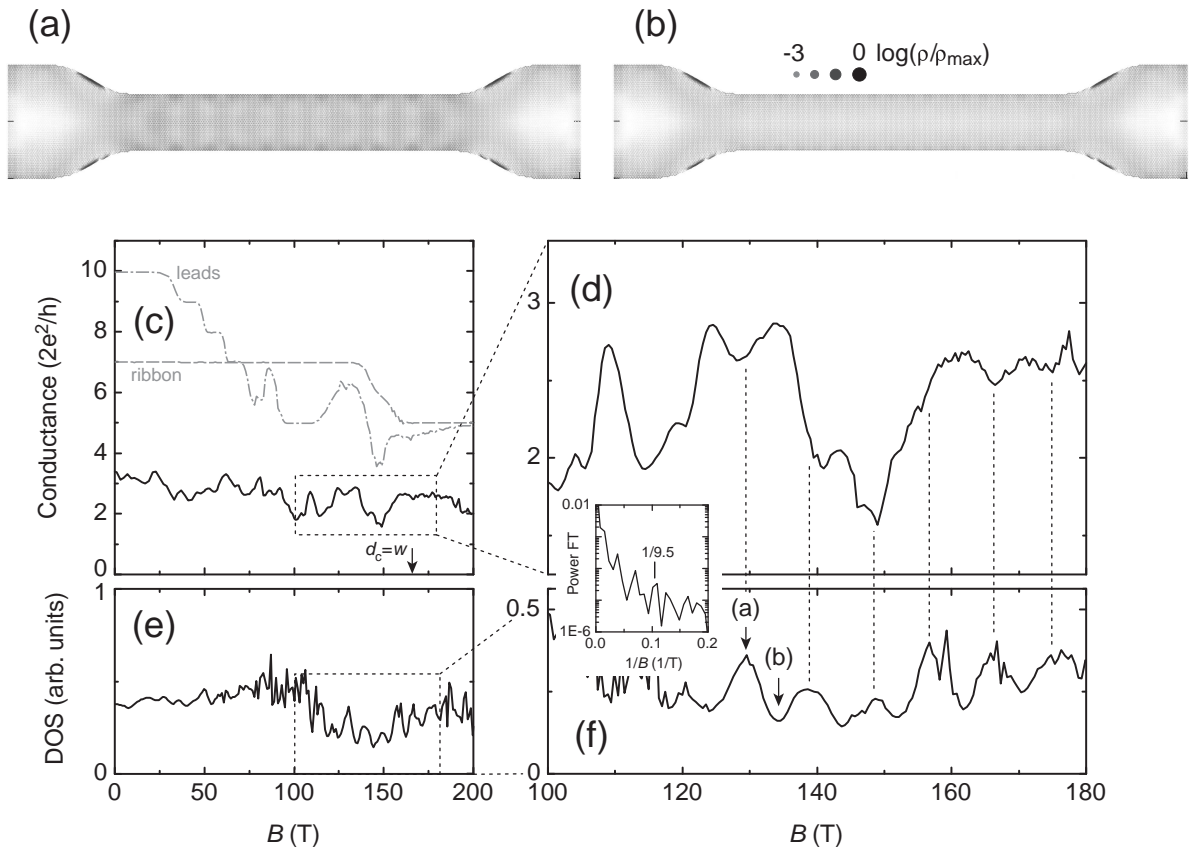}
\caption{(a)-(f) same as Fig. \ref{fig:1} but calculated using the Hartree approach. The device geometries used for both calculations are exactly the same; the geometry can be traced in the top plots (a) and (b). The inset between (d) and (f) shows power spectrum estimation for conductances using Fourier transform (FT), where a peak corresponding to $\Delta B=9.5$ T is marked.\cite{FT} The voltage on the gate electrode \cite{Ihn12, Ihn21} is 30 V. $T=20$ K.}
\label{fig:2}
\end{figure*}
%********************************************************* 

The long-range Coulomb interaction has several pronounced effects on the energy structure and transport properties of GNRs. First, it is known to cause charge accumulation at graphene boundaries.\cite{Sil08, Ihn12, Fer07, Shy10} This charge accumulation is related to formation of the triangular potential wells, which, in the magnetic field, accommodate extra forward and backward moving edge channels. These edge channels, according to the Landauer formula,\cite{Lan57} contribute to ballistic conduction\cite{Sil08} and thus causes $G$ to deviate from stair-like dependence predicted by the noninteracting theory, compare Figs. \ref{fig:1}(c) and \ref{fig:2}(c) for the leads and ribbon.\cite{Shy10, Car18} (Note that the self-consistent calculations in the Hartree theory are performed at finite temperature $T=20$ K, which in itself contributes as an averaging factor on the observable qualities, so that $G$ appears smoothed in Fig. \ref{fig:2}(c). In contrast, the noninteracting calculations are performed at $T=0$ K.) In the quantum Hall regime, the part of the structures that is away from the boundaries is filled out by the compressible strip,\cite{Shk92} which is an area with magneto-subband pinned to the Fermi energy.\cite{Shy10} The pinning effect reflects the screening ability of the system, in which free electrons can redistribute to minimize electrostatic energy, a property that is peculiar to a metallic system as opposite to an insulator. Another effect of Coulomb interaction is the strong electron localization along the interface physical boundaries (while expelling an electric current toward interior),\cite{Ihn21} compare local DOS in Figs. \ref{fig:1}(a),(b) and \ref{fig:2}(a),(b).

An important effect introduced by Coulomb interaction on transport in GNR is \textit{periodic} conductance oscillations at moderate $B$, see Figs. \ref{fig:2}(c),(d). Period estimation gives $\Delta B\approx 9.5$ T, which is corroborated by one of the peaks in the power factor spectrum of the Fourier transform.\cite{FT} Substitution $\Delta B$ into Eq. \eqref{eq:AB} gives $S=440$ nm$^2$. If adjusted by the finite decay length of the wave function, this agrees reasonably well with geometrical ribbon area $600$ nm$^2$. As in the non-interacting theory, $G$ dips match DOS peaks. This implies a resonant backscattering process due to the single-particle state formed in between of narrow-wide interfaces. This is supported by local DOS, $\rho$, visualization in Figs. \ref{fig:2}(a),(b), where the ribbon is uniformly filled in by the resonant state at $\rho$ peak and is "empty" otherwise. The structure of the resonant states is the same for all peaks. Contrary to the non-interacting theory, \textit{conductance oscillations in the Hartree approximation are due to the Aharonov-Bohm effect.} AB interference can be understood as a result of electron screening when energy levels of the resonant states inside the ribbon, which are poorly coupled to continuum of states in the leads, are adjusted to minimize the electrostatic energy of GNR. Every time a magnetic flux enclosed by the ribbon changes by the flux quantum, one single-particle state sweeps through $E_F$ and $G$ develops one full oscillation. AB oscillations, both in G and DOS, are more pronounced at intermediate magnetic fields, at which the edge channels start to form but the system is not yet deep in the quantum Hall regime. This condition might be expressed as $d_c\approx w$, where the cyclotron diameter $d_c$ in the self-consistent Hartree calculation is convenient to express via the charge density $n$: $d_c=2\hbar\sqrt{\pi n} /eB$. In the ribbon center, $n\approx5\times10^{17}$ m$^{-2}$, and for this $n$, the equality $d_c=w$ is marked by the arrow in Fig. \ref{fig:2}(f).

Regarding the name 'Aharonov-Bohm', it should be mentioned that Aharonov and Bohm originally proposed an experiment to show that observable effects could result from vector potential $\mathbf{A}$ on phases to the probability amplitudes associated with various electron paths, even though $B=0$ in those paths.\cite{Aha59} Here, no well-defined paths exist because $B$ is not sufficiently high to bring the system into the quantum Hall effect regime. However, the name 'Aharonov-Bohm' is still adopted, as is typically done in the solid-state experiments going under this name, because of similarity to the original AB effect in the sense of $G$ invariance under changing $B$ piercing the ribbon by one flux quanta.\cite{Datta}

Within the Hartree approximation, an analogous effect of correlation of the single-particle states at $E_F$, when the states depopulate sequentially in $B$, was shown to occur in an antidot-based Aharonov-Bohm interferometer in a conventional 2D electron gas.\cite{Ihn09-AD} Sequential depopulation implies that the number of charges in the region where quasi-localized states reside will change nonlinearly, which in turn indicates the possibility of Coulomb blockade physics. The Hartree theory, however, is unable to capture Coulomb blockade,\cite{Ihn09-AD} and the question whether an ideal GNR inherently supports Coulomb blockade remains open, which is in contrast to disordered GNRs where it was shown to be the dominant mechanism of charge transport.\cite{Mol09-Mol10, Bis15} %\cite{H-CB}

The Aharonov-Bohm conductance oscillations that are predicted here for GNRs have similarities with an experiment by Loosdrecht et al.\cite{Loo88} on quantum point contact in GaAs heterostructure, where two-terminal magnetoresistance revealed periodic in $B$ oscillations between the quantum Hall plateaux. Loosdrecht et al.\cite{Loo88} explained this as quantum interference due to tunneling between edge states across the point contact at the potential step at the entrance and exit of the constriction. However, apart from apparently the structure (which is long graphene ribbon here and short point contact in GaAs in Ref. \onlinecite{Loo88}), the difference is that in the present work magnetic fields are low for the quantum Hall plateaux to occur, and GNR is ideal without any imposed potentials.

\section{Conclusion}

Quantum-mechanical calculations suggest that ideal GNR in a geometry with wide leads, which is typically realized in experiments based on nanolithography,\cite{Bis15} should exhibit Fabry-P\'{e}rot or Aharonov-Bohm interference pattern in magnetoconductance, depending on whether or not Coulomb interactions dominate. This is a counter-intuitive finding because GNR is inherently open, atomically ideal, and contains no potential barriers. Electron quantum interference is related to strong electron scattering on the interfaces that connect the ribbon with electrodes and is a result of resonant backscattering via the quasi-bound state formed inside the ribbon. The structure of the quasi-bound states is different in noninteracting and Hartree theories; it might be accessed in local density of states spectroscopy measurements and further used to characterize interference effects in GNRs.

It is expected that above results should be valid for GNRs irrespective of edge terminations because the edge specifics are of less importance at operational magnetic fields and device widths. In addition, in GNRs of the same geometries as those considered here, the transport gaps have been shown to be nearly identical for armchair and zigzag terminations.\cite{Ihn21} 

It is also expected that above results should be applicable for GNRs in straight geometry provided by the strong electron scattering on potential barriers at metal-graphene contacts.\cite{Geh16, Ric13, Oks14}  

The defects should destroy the interference effects predicted in this study. However, these effects should become experimentally observable as the quality of the GNR samples improves.

The predicted interference effects hardly validate an ideal GNR for application as an interferometer (or etalon), mainly because the conductance oscillations are superimposed on relatively high background conductance and the visibility\cite{Wei17, Dep21-Ron21, Ji03} of such an interferometer would be low. One route to improve visibility might be to decrease the total number of electron quantization subbands. Another would be to design the proper shape and size of the interface regions, which are the sources of strong electron scattering in graphene.

\section{Acknowledgement}
This work was supported by SNIC 2021/22-961.

\twocolumngrid


\begin{thebibliography}{99}
\bibitem{FT} The Fourier transform was performed on conductances in the range $B=50\dots 177.5$ T using the rectangular window method. Averaging over three power spectrums, obtained for GNRs with slightly different interface lengths of 28, 29 and 30 armchair unit cells,\cite{Cas09, Ihn21} was applied for statistically more accurate result.
\end{thebibliography}
\end{document}